# The pioneering scientific endeavor and contributions of José María González Benito (1843-1903), the first Colombian modern astronomer

# Los esfuerzos y aportes científicos de José María González Benito (1843-1903), el primer astrónomo moderno colombiano

Freddy Moreno-Cárdenas[1], Santiago Vargas-Domínguez[2], Jorge Cuéllar-Moyano[1]

[1]Centro de Estudios Astrofísicos (CEAF), Gimnasio Campestre, Colombia.

[2]Universidad Nacional de Colombia, Observatorio Astronómico Nacional, Colombia.

**Abstract**

Astronomical interest within the current Colombian territory has its roots in the Botanical Expedition of the New Kingdom of Granada (1783-1813), which stimulated the creation of an astronomical observatory in 1803, the first one established in the New World to pursue systematic observations and meteorological studies. After the death in 1816 of its first director, Francisco José de Caldas (1768-1816), during the convulsive independence period, no major astronomical observations were made for decades, with few exceptions. In this work we delve into the contributions of the astronomer José María González Benito (1843-1903), main reactivator of the Observatorio Astronómico Nacional de Colombia in the second half of the 19th century, pointing out his pioneering efforts that put worldwide attention to it, and to his own private observatory making him one of the most committed figures to the development of astronomical sciences in the country and the most renowned Colombian in the international astronomical research scene of his time.

**Keywords**

José María González Benito; Observatorio Astronómico Nacional de Colombia; Flammarion Observatory; 19th century; comets; Mars.

**Resumen**

El interés astronómico dentro del actual territorio colombiano tiene sus raíces en la Expedición Botánica del Nuevo Reino de Granada (1783-1813), que impulsó la creación de un observatorio astronómico en 1803, el primero establecido en el Nuevo Mundo para realizar observaciones sistemáticas y estudios meteorológicos. Tras la muerte en 1816 de su primer director, Francisco José de Caldas (1768-1816), a causa del convulso período independentista, no se realizaron grandes observaciones astronómicas durante décadas, salvo contadas excepciones. En este trabajo profundizamos en los aportes del astrónomo colombiano José María González Benito (1843-1903), principal reactivador del Observatorio Astronómico Nacional de Colombia en la segunda mitad del siglo XIX, señalando sus esfuerzos pioneros que pusieron la atención mundial sobre esta instalación astronómica, y su propio observatorio privado. González Benito se destaca como una de las figuras más comprometidas con el desarrollo de la astronomía en el país, dando un impulso a un campo mayoritariamente considerado en el contexto histórico local como una herramienta práctica para tareas de ingeniería, como la cartografía. Argumentamos las razones para considerar a González Benito como el colombiano más destacado en el panorama de la investigación astronómica internacional de su época.

# 1. Introduction

In the second half of the 19th century, the figure of Colombian astronomer José Maria González Benito (1843-1903), hereafter JMGB, is inescapably connected with a large number of works of paramount relevance for the advancement of national science. His great ability to get involved in ideas and projects of various kinds is reflected in a long list of contributions that transcended purely scientific domains and transformed the social and cultural environment of his time. Besides the noteworthy scientific contributions of JMGB in astronomy, which are the main focus of this work, his pioneering contributions in other scientific areas should be considered. In 1871, JMGB was a pioneer in the teaching of stratigraphy, in the geology and paleontology courses that he led at the Escuela de Ciencias Naturales (School of Natural Sciences) in Universidad Nacional de Colombia. His curiosity also led his interests on bacteriology and microphotography. Since 1895, JMGB alternated lecturing this subject and solar physics at the Instituto Politécnico, an initiative that arose in a meeting with several academics at JMGB's house in 1893, with the aim of establishing a scientific center that would support the study of the new branches of human knowledge in the country; later this private institution would become Instituto Colombia. His studies on the microcosm are currently lost. However, it can be established that he used state-of-the-art imported microscopes, which allowed him to undertake studies of great practical utility, and which he communicated to the local Academia de Medicina (Academy of Medicine) formerly known as Sociedad de Medicina y Ciencias Naturales de Bogotá, standing as a national leader in this branch of knowledge.

The present work is the first using the information from JMGB's autobiography, a manuscript lost for more than a century that was accidentally found in 2018 and transcribed by Armando Martínez Garnica and Ramón García Piment (**González,** 2018). This material is complemented by the biographical description written by **Sánchez** (1906) and **Arias de Greiff** (1993), among other multiple references and information from national and international historical manuscripts and other sources.

Focusing on JMGB's passion for the cosmos, and after reviewing the material revealing his motivations and contributions to solving decisive problems in astronomy at his time, his expertise with instrumentation and countless efforts to get involved in fundamental astronomy research, we will consider JMGB as the first Colombian modern astronomer — what we call now an astrophysicist with vast technical and theoretical knowledge —, as it has also been appraised by additional present day astronomers (**Portilla**, 2017). Added to his purely scientific interest, JMGB was well aware of the importance of popularization of science to general audiences, as he undertook unnumbered efforts to spread useful knowledge to people, as part of Sociedad de la Luz (Society of Light) of the Instituto de Artes y Oficios (Institute of Arts and Crafts) formalized in 1872 in Bogotá, where he spread useful knowledge for the country, mainly writing about astronomical phenomena in newspapers of massive distribution, e.g. *La Ilustración*, and by motivating others to do so.

Historically, JMGB is mainly recognized for being the director of the Observatorio Astronómico Nacional de Colombia (OAN) for several periods, between 1868 and 1891. During his terms as director, this institution fulfilled for the first time the functions for which it was created, i.e. to carry out continuous astronomical observations with scientific rigor. JMGB did not hesitate to spend his own capital to adequate the premises of the OAN and to acquire sophisticated instrumentation, clear evidence of his commitment to the development of Colombian astronomy.

The OAN was founded in 1803 in Bogotá (Figure 1S), in the framework of the Expedición Botánica del Nuevo Reino de Granada (Botanical Expedition in New Granada) led by the Spanish physician José Celestino Mutis (1732-1808) and was mainly devoted to meteorological studies and astronomical observations related to cartography, in the first decades of the 19th century (**Bateman**, 1954; **Arias de Greiff**, 1993). Its first director was Francisco José de Caldas (1768-1816), who was shot by the royalists in 1816, during the fights for independence from the Spanish Kingdom (**Portilla**, 2020). No major astronomical observations were made in the following years, with a few exceptions, due to the large number of adverse political and social circumstances which surrounded the observatory and limited its administration and use (**Uricoechea**, 1860; **Torres Sánchez y Salazar Hurtado**, 2002).

In this investigation, we rescued and put together some of the most remarkable and innovative actions pursued by JMGB, including his role as director of the OAN, being its main reactivator in the second half of the 19th century, and highlighting the personal astronomy projects that he carried out in his private observatory.

**2. The scientific life of José María González Benito**

The Colombian astronomer JMGB (Figure 1) was born on September 1st 1843 in Zipaquirá, a town located 50 kilometers north of Bogotá in Colombia. From an early age, he was involved in drawing maps and surveying large salt mines around his place of birth, under the supervision of Manuel Ponce de León (1829-1899), one of the founders of the Sociedad Colombiana de Ingenieros (Colombian Society of Engineers), who served as his first prominent mentor, giving him private lessons (**Sánchez**, 1906; **Torres Sánchez y Salazar Hurtado**, 2002). From him, JMGB learned integral and differential calculus, physics, among other topics in science, at a time when education was difficult to access, since formal education was not standardized in the country. Afterwards, he became a young assistant of Indalecio Liévano (1834-1913), a distinguished Colombian engineer, and was able to contribute to the layout of the railway from Zipaquirá to Nemocón, including Sesquilé and Tausa (**Liévano**, 1875; **Tisnés Jiménez**, 1956; **González**, 2018). Many trips to rural areas foster the interest of JMGB in geology and paleontology, for which he decided to travel through the mountainous region from Sumapaz to Tunja (**Arias de Greiff**, 1993). His interest in astronomy arose in 1862 when Liévano decided to appoint him as his assistant in the management of the OAN, where he had been designated as director.

In 1864 JMGB traveled to Europe, where he enrolled at the Central School of Paris and attended some courses at Sorbonne University, which allowed him to meet the famous French scientists Urbain Jean Joseph Le Verrier (1811-1877) and Jean-Baptiste Boussingault (1801-1887), among others, who further catalyzed the enthusiasm for astronomy of the young Colombian (**Sánchez**, 1906). After returning to Colombia in 1866, JMGB was appointed by the government as assistant of the Oficina Central del Cuerpo de Ingenieros (Central Office of the Brigade of National Engineers), and also in charge of the diary of observations at the Observatorio Astronómico de Santafé de Bogotá (Astronomical Observatory of Santafé de Bogotá), since both places were united at that time. Two years later, JMGB received the position of teacher of astronomy and meteorology of the recently founded (1867) Universidad Nacional de Colombia (UNAL), being the first in charge of teaching these topics at university level in the country. At the same time, JMGB was appointed by Manuel Ancízar (1812-1882), provost of the university, as the new director of the OAN. However, JMGB would not last long in that position, since he decided to return to Zipaquirá, his hometown, to complete his previous and unfinished tasks, such as meteorological records from the location, and most importantly the Carta Geológica de la Sabana de Bogotá (Geological Chart of the Plateau of Bogotá), shown in Figure 2S, that represented eight years of intense work (**Arias de Greiff**, 1993). With this work JMGB obtains the honorable mention in the outstanding Exhibición de la Industria Nacional (Exhibition of the National Industry) of 1871, the most important in the country (**Revista Científica e Industrial**, 1871). Once finished, this work was published in scientific journals in Colombia and Germany, and it was used to establish new carbon mines in the region. In 1871 he returned to the UNAL, this time as a professor of geology and paleontology, being the founder in the country in the teaching of these topics, for which he used the collection of more than 5.000 pieces of crystallography, mineralogy, geology and paleontology samples, that he personally acquired in Paris, besides other materials collected in his exploration trips. Three months later he was notified about the designation, for the third time, as director of the OAN and in charge of the master classes in astronomy and meteorology at the Escuela de Ingeniería of the UNAL. The meteorological observations carried out were published in the Anales de la Universidad (Annals of the University) (1882a, 1882b, 1882c, 1882d, 1882e). In 1871, the creation of the Academia de Ciencias Naturales (Academy of Natural Sciences) was established by national decree to be integrated to the Escuela de Ciencias Naturales of the UNAL, with JMGB one of the members, together with other instructors of the university. During the same period, JMGB was invited to be a member of the Amigos de la Luz (Friends of the Light), a new society aimed to promote the popularization of science, through courses and outreach activities addressed to the general public, where he was mainly in charge of the topics related to geography and astronomy (**Sánchez**, 1906).

As for his academic work at the OAN, he resigned in 1872, due to differences with the directives and colleagues from the university regarding his duties, mainly because it was not accepted that JMGB refused the salary and that the expenses for utilities, cleaning, and conservation of the building and garden of the OAN were paid from his own resources. However, he would return for the fourth time to be director of the OAN in September the same year, but this time teaching classes as a professor of astronomy and geodesy in the School of Engineering. Once again, JMGB decided to spend some time in Europe, mainly in France and England, a period in which his position at the OAN was covered by Luis Maria Lleras Triana (1842-1885), a renowned mathematician and engineer (**Arias de Greiff**, 1993). During this trip, JMGB was able to visit several European observatories, such as the ones in Saint Petersburg and Moscow (**Sánchez**, 1906). In England, the Queen approved JMGB as Consul at Southampton for the United States of Colombia (**Bulletins and Other State Intelligence**, 1874).

In Paris, JMGB's interest to deepen his scientific knowledge enrolled him in an astronomy course given by the French astronomer Pierre Puiseux (1855-1928) and one in geology by the French geologist and mineralogist Gabriel-Auguste Daubrée (1814-1896), leading scientists of his generation, and initiate relations with European scientific institutions such as the Paris Observatory. In 1875, JMGB returned to Colombia, soon after receiving the membership of the Royal Astronomical Society in London. In 1882 JMGB was presented to the French Astronomical Society, forming part of it as a founding member. His last days were devoted to the creation of the Institute of Colombia, where the academies of mathematics, moral and ethical sciences, and social sciences would take part. Unfortunately, JMGB died the day before the inauguration of the Institute that was planned to be celebrated at the most important theater in Bogotá, on July 28, 1903 (**Sánchez**, 1906; **Arias de Greiff**, 1993).

### 3. His role as director of the Observatorio Astronómico Nacional de Colombia (OAN)

One of the first actions of JMGB in the OAN was the reconstruction of the original meridian spoiled during national civil wars, and in which he originally participated at an early age while being assistant of Indalecio Liévano in 1862 (**Arias** et al., 1987). Since then, JMGB was much involved in various actions that strengthened the development of the observatory infrastructure and astronomical instruments during the following half-century. Furthermore, JMGB stood out for making the OAN known in Europe, since his first trips to study in London and Paris. He established special connections in France with the astronomer Camille Flammarion (1842-1925), a major source of inspiration and information concerning astronomical issues, but also a close friend for the rest of his life. Flammarion highlighted many times that the OAN was located in a privileged place since it was the closest observatory to the equator, and this generates a very important advantage, such as the possibility of seeing the stars of both the northern and the Southern hemisphere. Another advantage was related to the altitude of more than 2600 m.a.s.l., defining it as one of the two highest observatories in the world, which is ideal for better imaging. These qualities were published in the journal of the French Astronomical Society, specifically in the list and description of the observatories of the world, published in volume VIII of the "Studies and readings on astronomy" with the statement "the Observatory of Bogotá is the closest to Ecuador and the highest in the world" (**Flammarion**, 1882).

Once in charge of the OAN, and after his trips to Europe, he identified which were the main requirements for an observatory. During the presidency of Rafael Nuñez (1825-1894), JMGB was able to receive a generous financial budget to acquire new instrumentation and carried out some refurbishments in the premises which included the improvement of the area surrounding the observatory with a beautiful garden (**Ibáñez**, 1891). Between 1868 to 1892, during his multiple stays at the head of the institution, JMGB transformed the OAN from a place lacking instrumentation (even a telescope) to an observatory with modern equipment, a new dome, a library and international recognition among the scientific community (**Quintero**, 2002).

A memorial plaque (Figure 3S) was installed by JMGB on 20th July 1881 (during the 71st commemoration of the independence of the country on the 20th of July 1810) in one of the walls at the OAN, to acknowledge the impulse given by president Núñez, allowing the reinstatement of astronomical observations in September 1880 thanks to the new instrumentation.

The main objectives to be carried out by the OAN proposed by JMGB (**González**, 1882d) and reflected in his statement: "Colombia would render a great service to science if the following works were carried out in this Observatory, which is considered of great importance by the scientific world", summarized as follows:

- A catalog of the stars of the southern hemisphere.
- A catalog of double and multiple stars in the same hemisphere.
- A catalog of the nebulae and stellar groups in the same region.
- A continuous study on asteroids.
- A special study of solar physics.
- Applications of spectral analysis to the study of celestial bodies.
- A sustained study on the zodiacal light.
- Special selenographic studies.
- Assiduous observations on the physical constitution of the planets.
- Application of photography to the study of the physical constitution of the Sun, the Moon, and the planets.
- A sustained study on shooting stars.
- Observation of the transit of Venus in front of the solar disk.

The main improvements that JMGB made to the OAN during the periods that he served as director, including actions and new duties, are summarized as follows:

- Construction of a movable dome.

- Ordinary meteorological observations.

- Acquisition of imported meteorology and astronomy equipment, e.g., an anemometer, hypsometer, spectrometers, and telescope, to be able to properly fulfill the different functions.

- Continuous communication with European observatories in England, France, Italy, and the Vatican; and in America, e.g., Chapultepec Observatory, despite difficulties in communication with Bogotá, and even in Africa, e.g., Algiers Observatory, (**González**, 1882c).

- As a consequence of the previous point, the OAN received numerous publications from the aforementioned observatories, keeping updated the local library (**González**, 1882b).

- Created an own publication of the OAN, which was entitled Anales del Observatorio Astronómico (Annals of the Astronomical Observatory), from which he produced six volumes between March and November in 1882.

- At the end of the 19th century, and due to the advance of communications, the need arose to establish a zero meridian and create time zones, for which the OAN was invited to participate in the 1881 meeting on the adoption of the prime meridian made in Washington. The astronomer JMGB was unable to attend but delegated the Colombian participation to a North American astronomer (**González**, 1882a).

- JMGB served as an academic international peer reviewer, e.g. for the De Large 1895 article.

## 4. The Flammarion Observatory

During his travels through Europe, JMGB had the opportunity to look for astronomical instrumentation and once in Colombia he decided to equip a private observatory. In 1880 he received the equipment and inaugurated the so-called Observatorio Flammarion (Flammarion Observatory) in Zipaquirá, his birthplace, in honor of his friend and renowned French academic Camille Flammarion. The Flammarion Observatory housed a 1.65-meter focal length telescope, a five-prism spectroscope, and many more implements such as chronometers, thermometers, microscopes, meridian circles, among others (**Flammarion**, 1882). However, on September 3, 1881, JMGB was appointed director of the OAN in Bogotá, accepting the position and thus decided to move his observatory to the capital, specifically in the neighborhood of Los Mártires. In this location, JMGB organized a very impressive site for a private observatory with all the facilities, instrumentation, library, gallery, and even lodgment for assistant astronomers (Figure 2). In May 1882, the Flammarion Observatory was inaugurated with the attendance of numerous figures from the Colombian academy, as well as the ambassadors of France and Chile (Figure 4S, left) as it honored the Republic of France (**Flammarion**, 1882). Numerous French newspapers reported the news from Colombia, among them, were *Le Petit Journal* (Figure 4S, right), *Le Spectator*, *L'Avenir de Vichy*, *Journal de L´Orne*, and *Progrès*, that announced: "the inauguration of the Flammarion Observatory in the United States of Colombia, very close to the equator and at a great height of 2640 m.a.s.l."

Nevertheless, the observatory building was not the most suitable one, and JMGB began the construction of a whole new building in 1892, in which the observatory site was designed to be placed on the third and fourth floors. At the definitive location on 16th Street, the Flammarion Observatory, with a design characterized by a singular movable construction (rotating booth), JMGB was able to pursue diurnal and night observations, and spectroscopy.

Figure 3 (top image) shows a visualization of Flammarion Observatory as inferred from photographs of the location that we found after a long exploration of photographic archives, including aerial photographs (lower left image) and the last image featuring the Observatory, from the late 1960s, nearly a decade before the building was demolished (bottom right image). It should be said that this is the first time the aspect of the Flammarion Observatory is revealed in an academic manuscript, as we could not find any visual reference of the construction apart from a short mention and photograph in **Revista Semana** (1951). The construction must have been a very remarkable assembly on the roofs of the Bogotá city landscape as shown in photographs from the 1940s and 1950s (Figure 4). While the works were being completed, the Flammarion Observatory had a temporary headquarters not far from its final location. Three years later, a new larger-diameter equatorial telescope, manufactured by Secretan company in Paris, was installed. From this date and during almost a decade, until the death of JMGB in 1903, numerous astronomical observations were made, as will be commented in section 5, together with meteorological measurements including temperature values from 1874 to 1895 (**Sánchez**, 1906), that were directly requested by Flammarion to compare with European records as reported in a private communication by **Flammarion** (1895), (Figure 5S). The interest in the further development of observational astronomy in Colombia motivated JMGB to propose to the Royal Astronomical Society the construction of another observatory at an altitude of 3300 m.a.s.l. in the surroundings of Bogotá (Figure 6S), with the participation of the British government and a private Colombian contribution (**MNRAS**, 1874).

Flammarion Observatory was operated by his son-in-law, Manuel Laverde Liévano (Figure 5), but there are almost no information concerning the observations that were pursued at the facilities, apart from some eclipse observations, e.g., the annular solar eclipse observed in Bogotá on 7th March 1951 with an equatorial 16-cm telescope, with two meters in focal length (**Revista Semana**, 1951). The Observatory counted with a lift, one the first for a private building in the country. In the second half of the 20th century and beyond there are no references about the facilities; the building was demolished in the 1980s, and there was now written evidence of its existence until we searched for corroboration during this investigation.

## 5. Astronomical research and main scientific contributions

As previously mentioned in section 2, JMGB had a strong interest in physical sciences in general. During his stays at the OAN he compiled continuous records of the climate in Bogotá that included: maximum, minimum, and average temperatures, wind direction, cloud cover, amount of rain, day and night irradiance, and a number of meteors per hour (**González**, 1871). He was also very interested in geomagnetic and seismological studies and installed in this house in Zipaquirá the instruments for magnetic and seismographic measurements, i.e., a magnetic needle experiment to register terrestrial magnetic field variations and a seismograph, respectively. On the first days of June 1870, he noticed variations in the position of the magnetic needle preceding the occurrence of an earthquake on June 4 and reported the detection (**González**, 1871). Once at the OAN, he installed the magnetic needle experiment on top of the building for continuous registration of the magnetic field direction, data he accordingly annotated in his diary of astronomical-physical observations. On August 30, 1871, he started perceiving abnormal changes in the position of the magnetic needle, marking 0º6'10'' towards the East, which continued increasing the following days. That reminded him of the detection from the previous year from Zipaquirá and made him suspect of a possible seismic movement, which indeed occurred on September 7, 1871, with a duration of about 15 seconds, while the magnetic needle was deviated 0º10'50'' towards the East, as JMGB reported (**González**, 1871).

Among all scientific interests of JMGB astronomy occupied the first place. In the following subsections, we present the main astronomical observations, calculations, and investigations carried out by JMGB, whose most important references were found in international sources.

**5.1 Solar observations**

One of the interesting phenomena attracting the attention of observers occurs when a planet transits in front of the solar disc. A transit of Mercury occurred in 1881 and was described in **González** (1882b), as follows: "This phenomenon took place on November 7 of last year, at 5 hours, 19 minutes p.m., Bogotá average time, and it could be observed in this Observatory under convenient conditions despite the proximity of the Sun to the horizon".

The transit of Venus, which had long been so concerned by scientists, despite being less common, is easier to observe due to the larger size of Venus compared to Mercury. The astronomer JMGB presented the ephemeris in the *Annals of the Astronomical Observatory* **González** (1882c) with the date and hours in which the transit of Venus would occur on December 6, 1882 (Table 1S). Due to the possible poor meteorological conditions at the location of the OAN, he established an additional observing point in Bogotá at the Flammarion Observatory (**González**, 1882c). The transit of Venus was eventually observed in Bogotá with careful attention, both at the OAN and the Flammarion Observatory. Venus presented the appearance of a large cherry, standing out against a greenish background (**Sánchez**, 1906). Solar observations were a recurrent source of interest for JMGB. In 1894 one of his drawings of a large sunspot, observed and drawn by him in August 1893, as evidence of the maximum of the solar cycle number 13 (Figure 7S), was published in *L´Astronomie* (**Flammarion**, 1893). The caption of the corresponding image says: "Director of the Observatorio Flammarion Bogotá. Observed since its formation. Drawing of the great sunspot of August 1893". The description of the sunspot was also included in **Flammarion** (1894).

**5.2 Mars**

In 1894 occurred the opposition of Mars. JMGB received a letter from Flammarion requesting the drawings of Mars made by the Colombian astronomer from his observatory in Bogotá, which was afterward published in 1895. The quality of JMGB's work can be certified by these observations. At that time, JMGB was a well-known astronomer in Europe and kept continuous communication with his French friend and colleague. Indeed, JMGB made the observations from the Flammarion Observatory and sent 24 drawings, four of which were published in *La Planète Mars et ses conditions d'habitabilité* (**Flammarion**, 1909). Such drawings had been completely unknown by the Colombian academic community until now that we have found them (Figure 6). The quality of the images made the famous Italian astronomer Giovanni Schiaparelli (1835-1910) highlight and use the work of JMGB (**Flammarion**, 1909) (Supplementary information)

The report from JMGB was followed by the letter signed by Schiaparelli (Figure 8S) commenting on the Colombian astronomer's drawings (**Flammarion**, 1909): "These observations are particularly interesting, given the altitude of this observatory established on the equator (4°35'48" N). At the height of 2640 m, the atmosphere is very clear. Mr. Gonzalez is a careful and sincere observer. Of the 24 drawings that the wise founder of this equatorial establishment wanted to direct us, we chose four to be annexed here in our general documentation. Remarkably, the polar notch and the Main Sea (Lake Mœris) could be observed with the aid of a 108 mm. As for the decrease in the red coloration of the planet with its elevation in the sky, this may be due in part to an effect of our atmosphere that acts on the coloration of the Moon and the Sun, and in part to the objective of the lens, less achromatize, perhaps, by the blue and violet rays".

**5.3 Meteor showers and meteorites**

Since his first contact with astronomy, JMGB had been interested in meteor showers. He observed the Leonids at dawn on November 14, 1867, and together with Liévano, the director of the OAN, he organized the observation of the phenomenon. To get more precise data, they contacted the "serenos", name given to the night guards of the city before the police existed, and after explaining what was going to happen, they trained them to annotate the number and characteristics of the observed meteors. Some inhabitants of Bogotá, who were unaware of what was happening thought they were witnessing the end of the world, due to the impressive scene with countless meteors according to the own words of JMGB, "nothing can compare with the grandeur of the spectacle: at one thirty minutes in the morning, time of departure from the radiant point, some shooting stars were seen, and from two to five, the number was immense, it really looked like a gigantic artificial fire that, radiating from Regulus in the constellation of Leo, spread throughout the celestial vault" (**Sánchez**, 1906). These observations were made at the Flammarion Observatory in Zipaquirá and also in Bogotá.

A young JMGB reported a maximum of unknown shooting stars on the night of November 24, 1872, whose origin was the constellation of Leo. These observations were published in volume V of the *Studies and readings on astronomy* (Figure 9S) printed in Paris (**Flammarion**, 1874). His work on this topic served to verify Giovanni Schiaparelli's theory, in which the Italian astronomer established that the meteor showers were the result of cometary disaggregation (**Flammarion** 1874). Many years later, JMGB observed the Leonids of 1899, for which he prepared a work explaining their origin and motivated the community to observe the phenomenon and share the data with him (**Sánchez**, 1906).

Concerning meteorites, JMGB wrote a detailed report of the most important event in the history of Colombia, the Meteorite of Santa Rosa de Viterbo (**Flammarion** 1874), that had fallen in 1810 and was studied by **Boussingault and Rivero** (1823) and presented by Alexander von Humboldt (1769-1859) to the French scientific community the same year.

**5.4 Comets**

In general, JMGB was a virtuous and devoted comet observer, with the advantage of being located close to the equator and therefore able to see comets in both hemispheres (**Sánchez**, 1906). During 1880 he observed a large number of comets, all visible to the naked eye, including the great southern comet, also called 1880 I, the comet 1880 V discovered by Cooper which reached a magnitude close to 5, and the comet Hartwig (1880 III) of 5 to 6 magnitude (**Vsekhsvyatskii**, 1964) In 1881, he reported the observation of seven comets from Bogotá: Comet 1881 II discovered by the astronomer Lewis Swift (1820-1913) from Rochester, New York, comet 1881 III discovered by Tebbutt from Australia, the most beautiful according to JMGB and widely seen in Colombia, the periodical comet Encke, comet 1881 V Barnard, comet 1881 VI that was discovered by Denning, comet Schaeberle 1881 IV, and comet 1881 VIII, also discovered by Swift (**González**, 1882b).

While being director of the OAN, JMGB reported on another visitor, in the note he addressed to the Secretaría de Instrucción Pública (Secretary of Public Instruction) in Colombia on June 22, 1882 (**González**, 1882b), which reads as follows: "I have observed this afternoon at 6:30 p.m. a large comet located south of the planet Venus, not far from Procyon, where the constellations Cancer, Can Minor and Gemini border. Its core is extremely bright, it has the intensity of a second magnitude star, and its tail, a uniform matte white, extends more than 10º. A comet of such magnitude and beauty has not been observed for a long time. Its tail is directed to Sigma of the constellation of Hydra". This is known as the Comet of Wells (1882 I), named after the observer who discovered it on March 18 of that year from Albany (USA), in the constellation of Hercules (**Vsekhsvyatskii**, 1964).

During its passage, the comet 1882 I traveled the constellations of the Lyre, Cepheus, Dragon, Giraffe, Perseus, Auriga, and the Bull, until reaching their perihelion, on June 11 at a distance of 0.06 astronomical units (AU). Then, it went through Orion, Gemini, and Cancer. Several observatories reported seeing it very close to the Sun during the day. By June 17, its tail was about 40º long. On the 22nd of the same month, Pakl reported that its core was well defined and had a brightness equivalent to that of a 2nd magnitude star, with a 2.5º wide fan-shaped tail, characteristics similar to those observed by the JMGB and hardly repeatable since its period is more than a million years. Table 1 lists some of the parameters found by JMGB and another report from **Vsekhsvyatskii** (1964).

At the beginning of September 1882, one of the most striking comets of the 19th century was discovered and called the Great Comet of September (1882 II), due to the great brightness that reached in the middle of the month the equivalent to a star of 0 magnitude and the length of its tail reaches between 15º and 20º (in early October). The first reports of this comet originated in the Gulf of Guinea and the Cape of Good Hope, on September 1st (**Kronk**, 1999) and others from New Zealand from an Italian ship on the 1st and 3rd of the same month (**Vsekhsvyatskii**, 1964). Very concerned about what was happening in the sky, JMGB carried out astronomical observations during the early hours on 14 August and reported having seen the comet at dawn on that day (Figure 10S). Nonetheless, in the following days, he could not observe the comet due to poor weather conditions (**González**, 1882e), but received observing reports at the beginning of the same month, from Boyacá, a region a few hundred kilometers from Bogotá. With the acquired information, JMGB wrote to several European observatories the observation of comet 1882 II in mid-September, although in *L'Astronomie* only two very short reports from him appeared on October 5 and 20. Due to bronchitis that developed as a result of the observation on August 14, JMGB could not continue with the study of the orbital elements of the comet and delegated this work to his colleagues Benjamin Ferreira (1857-1918) and Eloy B. de Castro, who took the coordinates and they made drawings of the comet, with a particular star shape, from the Flammarion Observatory. It is unknown why Flammarion did not publish such observations of this very relevant comet, which apparently was seen in Colombia before other places in the world. From the data collected by JMGB, he established that the diameter of the comet's hair was 6' 55" and that of the nucleus of 2' 2" for the beginning of October 1882, much greater than what he measured in June when the hair was only one minute in diameter. Furthermore, JMGB calculated that for October 3, the comet would travel an angular distance of 4° 9´ 30" in 24 hours and would reach a speed of 30 leagues per second, that is, four times the speed of the Earth in its orbit around the Sun (**González**, 1882e).

In *L'Astronomie* (1893) the news of the discovery of a new comet by the French astronomer Ferdinand Quénisset (1872-1951) on July 9, were presented (**Flammarion**, 1893). Quénisset, who was working at the Juvisy Observatory founded by Flammarion the same year, telegraphed Félix Tisserand (1845-1896), director of the Paris Observatory, and the Central Office in Kiel, Germany, to communicate his discovery. The same journal included reports from observers in various parts of the world. Eyewitnesses in Minnewasta, New York, claimed to observe the magnificent comet the day before, in the constellation Lynx and having a tail extended to the pole star. The amateur astronomer Alfred A. Rordame (1862-1931), from Utah, also observed and reported the comet to Lewis Swift in Rochester. The one from Randolph Sperra, an observer from Massachusetts, from June 19, seems to be the first report (**Vsekhsvyatskii**, 1964) observing with the naked eye. In Colombia, JMGB reported having seen it from Bogotá on July 1 and the following days (**Sánchez**, 1906), and sent the ephemeris for the second semester in 1893, drawings of the comet and its orbital elements (Table 2).

It is unknown why the earlier report sent by JMGB, compared to the one from the official discoverer Quénisset, was not considered (**Arias de Greiff**, 1993), which perhaps would have represented changing the name from comet 1893 II Rordame-Quénisset to Sperra-González, and the possibility of having, once more, the first comet in history discovered by a Colombian.

An example of the intense observing activity developed by JMGB at his private observatory is evidenced in the manuscript written on September 14, 1898, with the following statement: "the comet that is currently visible is not only one, but there are no less than five in the following order. Comet Coddington, SE of Antares or Alpha of the Scorpion, visible from half-past six in the afternoon; very dim and to observe well, the telescope is necessary. Second, a telescopic one in the constellation Capricorn is nearly visible from the same time as the previous one. Third, Encke's comet, whose period is three and a quarter years, is visible to the naked eye at the moment, from two in the morning between Gemini and Canis Major. Fourth, another telescopic one located between the polar star and the Alpha of Perseus, in the constellation of the Giraffe, observed in Bogotá from eleven o'clock at night. And fifth, comet Wolf, seen in 1891 and slightly visible in the constellation of Aries, from ten at night, and passes through the meridian at three in the morning" (**Sánchez**, 1906).

**5.5 Stargazing**

The French Astronomical Society delegated to the OAN the task of methodical observation of the sky, and therefore entrusted to JMGB the study of astronomical phenomena located in declinations between 40°N and 55°N (**Sánchez,** 1906). Apparently, JMGB worked continuously on this project, according to an annotation found in **Sánchez** (1906), although no publications on the subject have been found yet. Furthermore, JMGB also received requests to solve doubts that the astronomy of the time presented, such as the request sent from the Italian astronomers Annibale de Gasparis (1819-1892), director of the Naples Observatory, and Giussepe Franchini, in 1882. Franchini had the hypothesis that the sky rotates in mass around the North Pole with a convergent movement, in such a way that the phenomena observed in the boreal hemisphere should be very different from those observed in the southern hemisphere; while de Gasparis maintained that there was uniformity in both regions. The reason for receiving this request was certainly due to the privileged observation of both hemispheres from the advantageous position near the equator at the OAN. The response from JMGB, after having observed stars in both hemispheres was: "that a star located at 80° southern declination, describes a parallel equal to that traversed by stars located at 80° northern declination. As for the apparent movement of the Milky Way, for an observer placed in the boreal hemisphere, the same is verified for the one who observes it in the austral region; but these are only appearances, since the movement is uniform as a whole, as is in particular that of each of the stars that compose it, the law is general" (**Sánchez**, 1906).

JMGB maintained continuous communication with Flammarion and was informed about the most important projects carried out by the French Astronomical Society, among which there was one he wanted to participate actively, the revision of the measurement of the meridian arc at the equator. This research was suggested by the renowned mathematician Henri Poincaré (1854-1912) and supported by the International Geodesic Association in 1889. Due to political issues, measurements could only start in 1899 right on the border between Colombia and Ecuador and extended through the latter to the border with Peru (**Littlehales**, 1907). With that on the agenda, JMGB trained a group of Colombian engineers and established communication with the French embassy in Bogotá, to bring to Colombia the necessary equipment to extend the measurement within the Colombian territory. Nevertheless, in 1899 the worst civil war that the Colombian nation has ever faced began, and the project could not be carried out as initially planned (**González**, 1902; **Schiavon y Rollet**, 2017).

## 6. Analysis of González Benito's pioneering work in Colombia

In the Colombia of the second half of the 20th century, the figure of JMGB is inescapably connected with a large number of works of great relevance for the advancement of science. His great ability to get involved in ideas and projects of various kinds from an early age is reflected in a long list of contributions that transcended purely scientific areas to even transform the social and cultural environment of his time. As an example, his family business "Gonzalez Benito Hermanos", which he ran with his brothers Eugenio and Fabián in the center of Bogotá (a few blocks from the OAN), had the first private telephone line in the city, in December 1884, to establish the connection with JMGB's home in the neighborhood of Chapinero, about seven kilometers away (**El Comercio**, 1884). It represented the very beginning of the public telephone network in the country.

Focusing now on JMGB's passion for the cosmos, and his role in modern astronomical problems of his time, e.g., investigations about Mars, origin of meteor showers, understanding of solar activity, comet observations, among others, there are adequate elements to consider JMGB as the first Colombian modern astronomer.

As director of the OAN and of his private observatory, JMGB made numerous observations on different astronomical topics, being reports on comets among the most extended ones, with the difficulties from the mostly cloudy conditions at this location. He observed three comets during 1880, seven in 1881, and comet 1882 II, two weeks before the first report of this comet from New Zealand made on September 1 in 1882. In 1893 he observed the comet Rordame-Quénisset, nine days before the observations made from the United States and France (designated as the official discovery). In 1898 JMGB identified five more comets. JMGB also observed meteor showers: the Leonids, on 14 November 1867 and 13 November 1899, and the unknown one he reported on 24 November 1872 in Andromeda.

Besides his scientific qualities, we must also highlight JMGB's virtues and commitment as an administrator. When appointed as director of the OAN, the building was in a state of abandonment mainly due to the difficult social and political conditions that Colombia had throughout the 19th century after its independence. He recovered the premises and refurbished part of the infrastructure and instrumentation and did not hesitate to spend his own capital to do so, clear evidence of his commitment to the development of Colombian astronomy. As no other Colombian did in the astronomy international scenario, even many decades after his death, JMGB had a great ability to establish and maintain relationships with relevant people within the world of science, in particular astronomy. Due to his education and his direct contact with the European astronomical environment, he identified and defined the functions that a modern observatory should fulfill. In Europe, he had the opportunity to acquire telescopes and other equipment and thus was able to supply both, his private Flammarion Observatory and the OAN, with state-of-the-art instrumentation.

The interest of JMGB to consolidate more observing sites in Colombia, motivated him to propose to scientific societies in the United Kingdom and France the construction of an observatory at an altitude of 3300 m.a.s.l., which would include the participation of the foreign governments and a private Colombian contribution to be funded. Unfortunately, the project did not succeed after JMGB's death, but his effort to consolidate such an observational facility in the high Colombian mountains shows a clear vision for the development of astronomy and the advantages that this would bring to scientific research in Colombia. Furthermore, his determining connections with some of the international inquiries on astronomy issues of his time, demonstrate his robust scientific profile and worldwide significance.

The Flammarion Observatory was his life project. JMGB and Flammarion kept always warm and continuous communication; the French astronomer was the best man at JMGB's wedding. JMGB created the Flammarion Scientific Society, which was the first in the world to bear this name, with many others formalized in following years based on the Colombian experience, such as in the cities of Jaen, Argentan, Marseille, and Bruxelles. In 1893, JMGB was presented by Flammarion and Anatole Bouquet de La Grey to the Astronomical Society of France as a founding member. His initiatives and dynamism allowed the first collaboration of Colombian astronomy to take place within an international network, exchanging information and knowledge.

While still immersed in civil wars and with poor interest and understanding of science by the society of his time, JMGB promote fundamental studies in astronomy showing that it was possible to make important contributions from Colombia, without extensive support from the academic community. An illustration of this is the fact that his detractors finally managed to prevent him from entering the OAN, claiming that the science the country needed should not give room for JMGB's interest in fundamental astronomy, mainly considered as useless and being far from the scientific development that was supposed to be entirely aligned with the premise of astronomy as a tool for engineering towards the development of the country. After JMGB was forbidden to enter the OAN he responded with the phrase "they do not realize that where there is an observer equipped with instruments, there is, in fact, an observatory" (**Sánchez**, 1906). While part of the national academic community seemed to be closing its doors to him, internationally JMGB continued to gain notoriety.

Flammarion, a prominent figure in astronomy of his time, had a notorious appreciation towards JMGB and included him in the novel "Fin du Monde" where JMGB is one of the protagonists featuring the Chancellor of the Colombian Academy of Sciences, and one of the world's notable scientists who attends a meeting in Paris to assess the damage that the Earth will suffer in the event of an imminent collision with a comet (**Flammarion**, 1894a) and wrote "Everyone knew that he was the founder of an observatory located on the same equinoctial line, three thousand meters high, from which the entire planet was dominated and both celestial poles were visible at the same time … His universal fame also contributed to his being heard with the utmost attention." The observations and drawings made by JMGB on the characteristics of the Martian surface helped Schiaparelli and other later astronomers to develop hypotheses about the possibility of the existence of life on the planet, which today manifests itself with the trips and sending of probes to the red planet, and the topics currently embraced by astrobiology.

During his academic life, JMGB was a member of numerous local societies and groups and distinguished with multiple honors. Among them: Academia de Ciencias Naturales (1871), Instituto de Artes y Oficios (1872), Sociedad Politécnica (1876), Ateneo de Bogotá (1884) and founder member of the Sociedad Colombiana de Ingenieros and the Instituto Colombia which was created under the same principles of the French Academy. He was also a member of international societies and institutions (**Revista Ilustrada**, 1898; **Arias de Greiff**, 1993), such as the Royal Astronomical Society (1875), British Science Association (1875), Society of Geographical and Historical Studies of Salvador (1892), Universal Academy of Arts and Sciences of Brussels (1892), French Astronomical Society (1898), Official of the French Academy (1898), Belgian Astronomical Society (1898), Astronomical Society of the Pacific (1890s), and French Public Instruction Officer (1903).

In the Bulletin of the French Astronomical Society in 1903 the obituary of JMGB was published soon after he passed away (Figure 7), where he is recognized as an "elite spirit and noble heart".

**7. Conclusions**

In this work we have collected evidence from well-known and other more undisclosed sources, supporting the fact that JMGB was one of the most important figures in the history of Colombian astronomy, mainly due to the advances and discoveries that he achieved as a result of innumerable investigations and projects. Although much of his history and works remain lost due to his sudden death. Despite his intense scientific life, the legacy of JMGB seems not to be properly acknowledged in Colombia, and his name diluted internationally, being some of the reasons that led us to embark on this work. JMGB's contributions to astronomy were mostly published in international journals, but we believe that most of his works are lost. The overall contributions, international collaborations and accomplishments made by JMGB to fundamental astronomy cannot be undervalued if considering the topics and questions that were haunting the heads of the astronomy community worldwide.

According to the evidence, the work of JMGB was not utterly very well appreciated by the academic community in Colombia at that time, that were more focused on the immediate application of astronomy for engineering tasks such as cartography, delimitation of the territory and issues related to the standardization of the local time used for telegraphic purposes and communications in general as recently investigated by **Quintero** (2002) y **Benavides** (2020). JMGB was not very much into those practical aspects but moved by fundamental astronomy issues, ideas that were not aligned with the expected development of the country as expressed by Abelardo Ramos (1852-1900), president of the Colombian Society of Engineers, in the editorial of the journal of the Society written a couple of years before JMGB was not renewed as director of the OAN (**Ramos**, 1890).

A work by the historian Camilo Quintero Toro in Purcell and Arias Trujillo (2014) suggests that the figure of JMGB was intentionally enshrouded by the one of Julio Garavito Armero (1865-1920). Garavito, the successor of JMGB as director of the OAN, is perhaps the most renowned Colombian astronomer of all time, in particular since 1970 when a crater on the Moon was named after him (**Arias et al.**, 1987). Nevertheless, at his time, Garavito was by far less known in the astronomical worldwide community than JMGB did, and never published his works in international journals. Garavito's duties as director of the OAN were mostly aligned with the engineering purposes defended by Ramos and other members of the academic community, diluting much of the astronomical work in Colombia that JMGB had led. During the first half of the twentieth century, astronomy in Colombia did not have significant changes that would resume the pioneering actions of JMGB.

With this work, we intend to spread the ideas, projects, and legacy of JMGB to a wider academic and general audience, after collecting remarkable chronological documentation (Table 2S) and evidence that remained unknown for more than a century. We hope that much more of JMGB's legacy can be recovered and to appropriately recognize the figure of the can be considered the father of Colombian modern astronomy.

**Acknowledgements**

**Contribution of the authors**

The contribution of the authors to the document is divided into bibliographic search and contextualization by JCM, and research, bibliographic search, analysis, and writing of the manuscript by FMC and SVD.

**References**

**Arias, J., Arboleda, L.C. Espinosa Baquero, A.** (1987). *Historia social de la ciencia en Colombia*. Colciencias, pp 92-105.

**Arias de Greiff, J.** (1993). *La astronomía en Colombia.* Academia Colombiana de Ciencias Exactas, Físicas y Naturales.

**Bateman, A.D.** (1954). *El Observatorio Astronómico de Bogotá*. Universidad Nacional de Colombia.

**Benavides P.** (2020) Pies en el cielo, ojos en la tierra. Modos de ordenar, tiempos y fronteras en la circulación de saberes astronómicos (1865-1902). Tesis doctoral. U. de los Andes.

**Boussingault, J.B., Rivero M.** (1823). Memoria sobre diferentes masas de hierro, encontradas en la Cordillera Oriental de los Andes. *Boletín de Historia y Antigüedades.* Volumen XXVIII, Número 31. Academia Colombiana de Historia, pp. 572-577.

**Bulletins and Other State Intelligence,** (1874). part 1, p. 238.

**El Comercio** (1884). El teléfono. 3 de diciembre 1884.

**Flammarion C.** (1874). *Etudes et Lectures sur L´Astronomie*. Tomo 5. Gauthier-Villars Imprimeur Libraire, pp. 170.

**Flammarion, C.** (1882). *L´Astronomie*. Societé Astronomique de France. Paris, p. 31.

**Flammarion, C.(**1893). *L´Astronomie*. Societé Astronomique de France. Paris 1893, pp. 104, 228

**Flammarion, C.** (1894a). *La fin du monde*, pp. 63, 64, 67.

**Flammarion, C.** (1909). *La Planete Mars et ses conditons d'habitabilité*. Tomo II. Gauthier-Villars Imprimeur Libraire, pp. 74-77

**González, J.** (1871). Física y Meteorología. Temblor del día 7 de septiembre de 1871. *Revista Científica e Industria*l. Bogotá, pp. 12,13.

**González, J.** (1874). *Proposed observatory at Bogotá, South America. Monthly Notices of the Royal Astronomical Society*, *35*(1), 12.

**González, J.(**1882a). *Anales del Observatorio Astronómico Nacional de Bogotá*. marzo de 1882.

**González, J.** (1882b). *Anales del Observatorio Nacional Astronómico de Bogotá*. junio de 1882

**González, J.** (1882c). *Anales del Observatorio Astronómico Nacional de Bogotá*, (6), 82.

**González, J.** (1882d). *Anales del Observatorio Astronómico Nacional de Bogotá*, (3), 82.

**González, J.** (1882e). *Anales del Observatorio Astronómico Nacional de Bogotá*, (5), 66.

**González, J.(**1882e). Fin del Mundo. *La Ilustración*, Bogotá. pp. 158.

**González J. M.** (1902). La mesure de l'arc de méridien a l'equateur. *Bulletin de la Société Astronomique de France*. Paris, 90-91.

**González, J.** (2018). Autobiografía de José María González, Transcrita por:

**Martínez Garnica, A. García Piment, R.** (año). Archivo General de la Nación *Revista Memoria,* (19), 62-104.

**Ibáñez, P. M. (**1891). Las Crónicas de Bogotá y sus inmediaciones. Imprenta de la luz.

**Kronk, G. W.** (1999). Cometography a catalog of comets 1800-1899.Volume 2

**Liévano, I. (**1875). Apéndice de las Investigaciones científicas publicadas en 1871 por Indalecio Liévano.

**Littlehales, G.W.** (1907). The Recent Scientific Missions for the Measurement of Arcs of the Meridian in Spitzbergen and Ecuador. *Bulletin of the American Geographical Society*, *39*(11), 641-653.

**Portilla, J.G**. (2017). Observatorio Astronómico Nacional: sobreviviente del primer intento de construcción de ciencia nacional. En E. Restrepo, C. H.
Sánchez y G. Silva (eds.), Patrimonio de la nación. Tomo I. Colección del sesquicentenario. Bogotá: Nomos Impresores.

**Portilla, J.G.** (2020). *Firmamento y atlas terrestre: la astronomía que practicó Francisco José de Caldas*. Facultad de Ciencias Sede Bogotá. Universidad Nacional de Colombia.

**Quintero, C.** (2002). La astronomía en Colombia, 1867-1949. *Documentos CESO*, 17, 1.

**Revista Científica e Industrial** (1871), p. 12

**Revista Ilustrada,** (1898). 3, agosto 4.

**Revista Semana,** (1951). Astronomía. Retina gigante. Revista Semana de 17 de marzo 1951, pp. 26-27.

**Sánchez, D.** (1906). Biografía de José María González Benito. *Anales de Ingeniería*, *XIV* (165-166).

**Schiavon, M., Rollet L.t (**dir.) (2017). *Pour une histoire du Bureau des Longitudes* (1795- 1932), PUN-Edulor, Nancy.

**Ramos, A.** (1890). Editorial. *Anales de Ingeniería,* 37, 7.

**Tisnés Jiménez, R. M.** 1956. *Capítulos de Historia Zipaquireña* (1480-1830), Vol 1. p. 678.

**Torres Sánchez J., Salazar Hurtado, L.A.** (2002). *Introducción a la Historia de la Ingeniería y la Educación en Colombia*. Universidad Nacional de Colombia.

**Uricoechea, E.** (1860). *Contribuciones de Colombia a las Ciencias i a las Artes, con la cooperación de la sociedad de naturalistas Neo-Granadinos.*

**Vsekhsvyatskii, S.K. (**1964). *Physical Characteristics of Comets*. NASA and National Science Foundation. Jerusalem. 1964.

**Figuras**

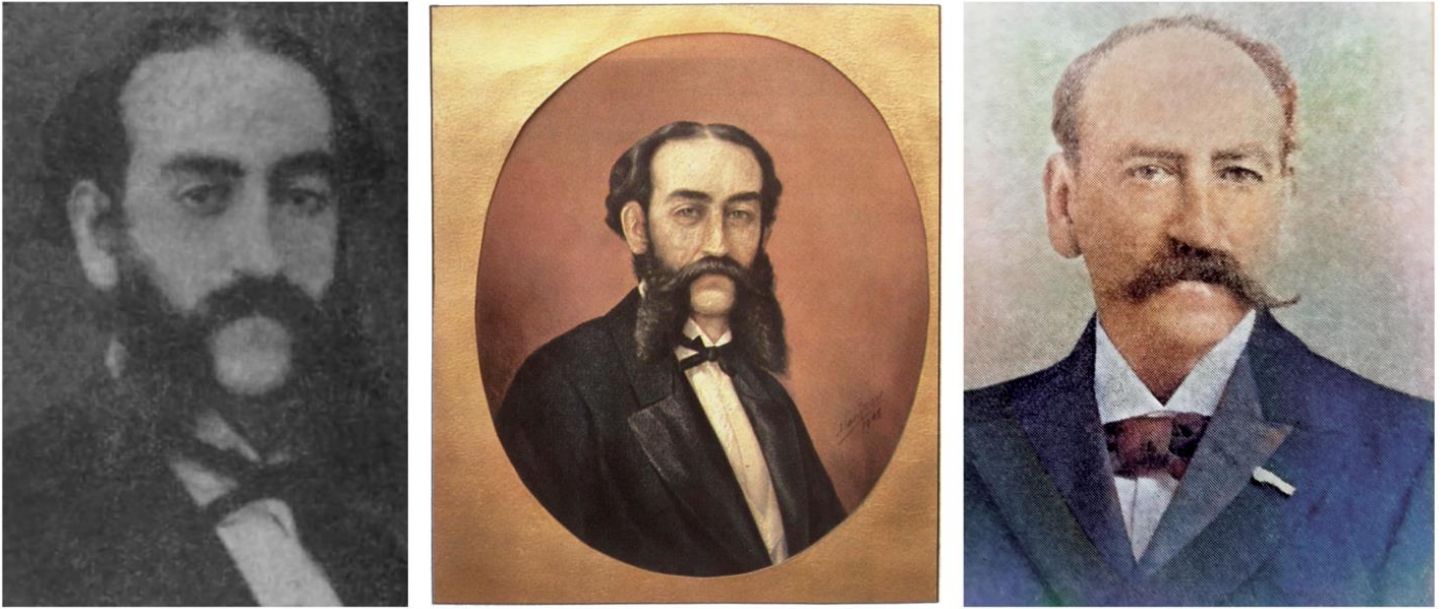

Figure 1: Portraits of José María González Benito. Left. Photograph of the Colombian astronomer in his 30s. Middle: Painting made in 1948 and located in the OAN (photo taken by the authors), which we argued was based on the photograph on the left. Right: Colorized photograph of the astronomer in his 50s (**Sánchez**, 1906).

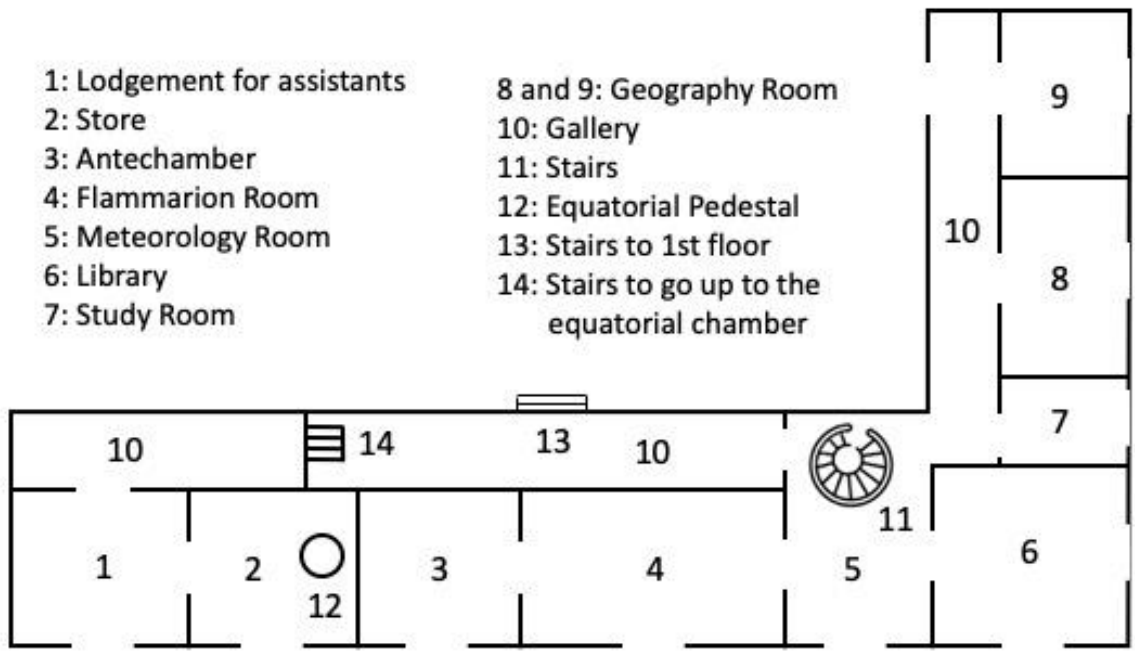

Figure 2: Sketch of the configuration of the Flammarion Observatory inaugurated in 1873 in Bogotá, as it was designed by JMGB. The image was made by the authors based on the notes and drawings of JMGB found in the repository of the Biblioteca Nacional de Colombia.

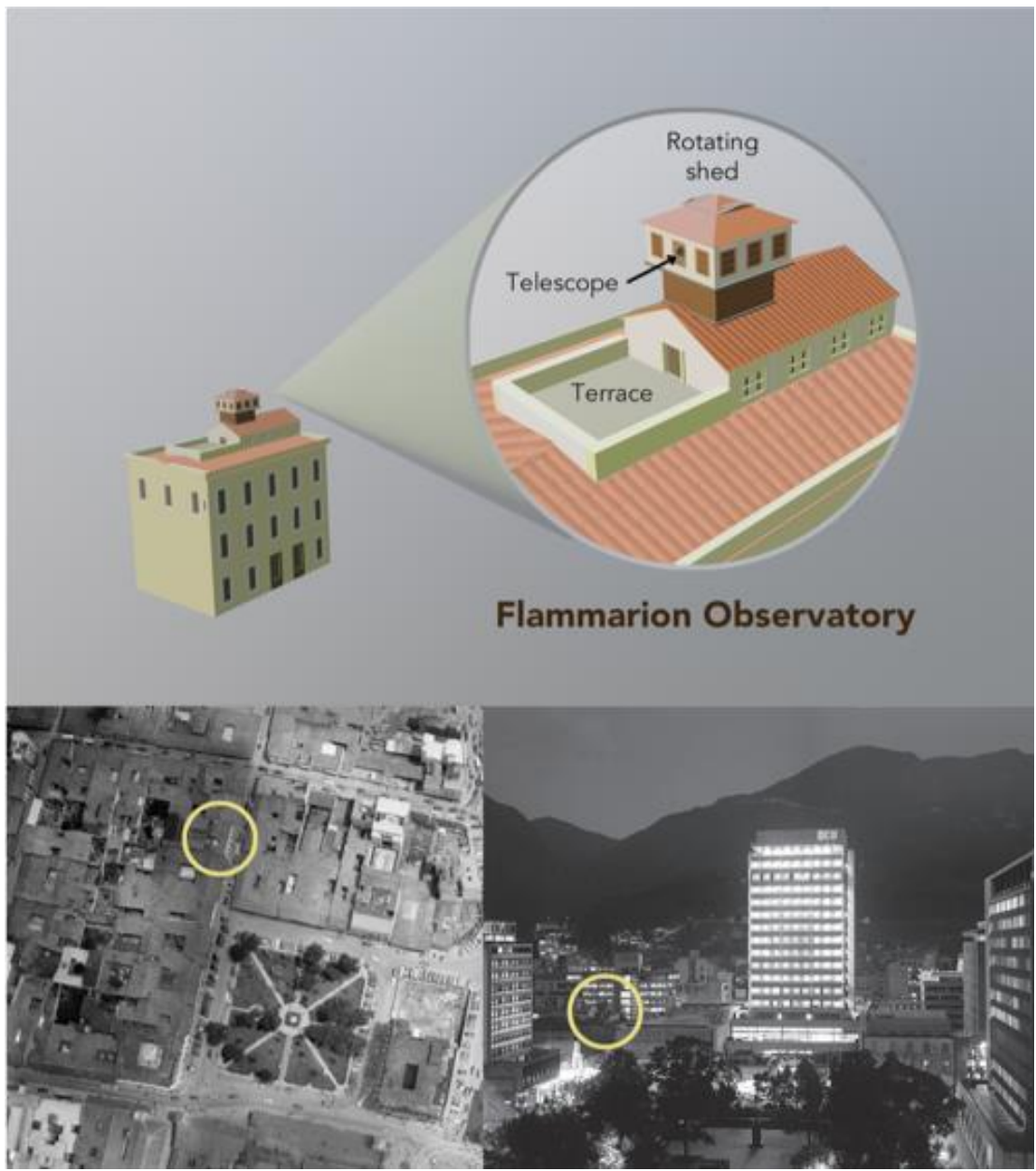

Figure 3: Flammarion Observatory. Top: 3D sketch of the house that supports the Flammarion Observatory on its terrace, made by the authors based on several historic photographs of Bogotá, showing the construction from different angles, including the aerial view from the 1940s (lower left image) and the last image in which the authors recognize the observatory, from the late 1960s. (lower right panel).

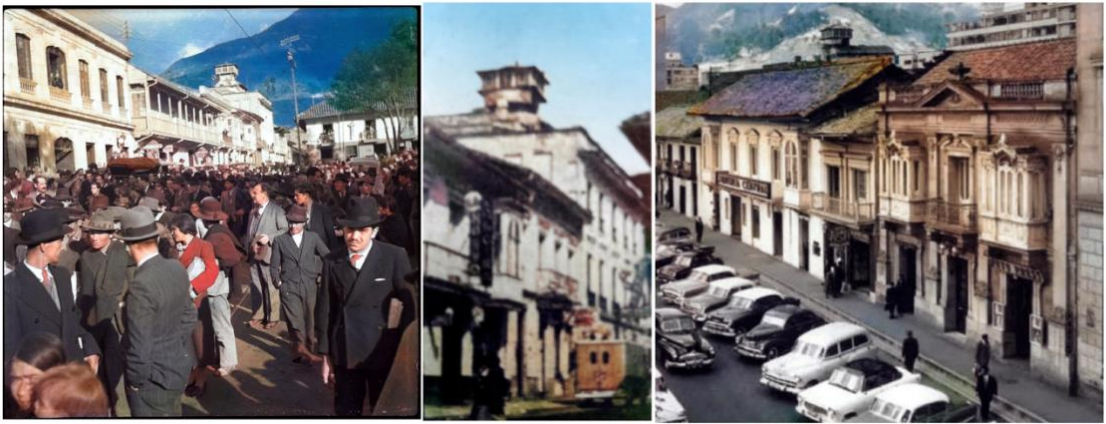

Figure 4. Photographs of the surroundings of the Flammarion Observatory where the construction of the booth stands out, raised about 20 meters above the ground. Left: Unpublished photograph of the funeral of Margarita Villaquirá ("crazy Margarita"; an iconic local figure), which took place in January 1942. Middle: Photograph from 1951 found in Revista Semana. Right: Photograph of the decade of 1940 with the Observatory Flammarion visible at the background.

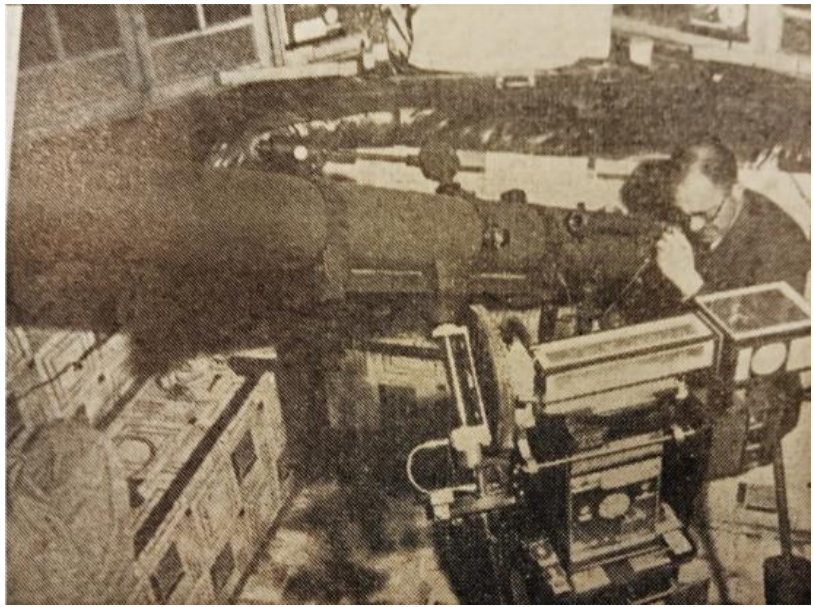

Figure 5: Photograph of the interior of the Flamarion Observatory as it was preserved by mid of the XX century The image shows Manuel Laverde Liévano, JMGB's son-in-law, making the observation with the main telescope and instrumentation (Revista Semana, 1951).

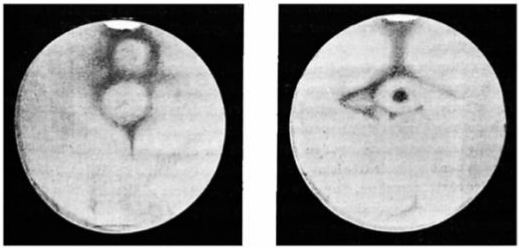
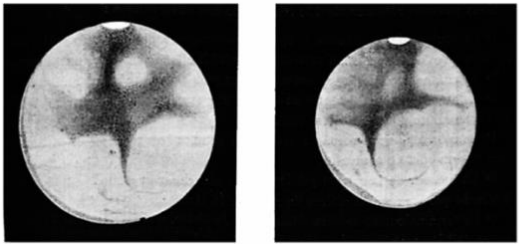
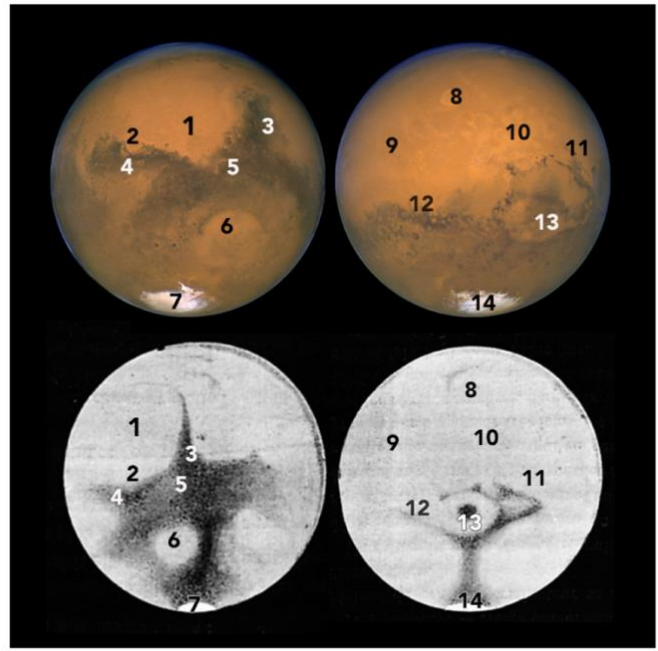

Figure 6: Mars observations. Left: From JMGB at the Flammarion Observatory, extracted from the publication La Planetè Mars et ces conditions D'habitabilité (Flammarion,1895). Right: Comparison of Mars observations acquired in 1999 by the Hubble Space Telescope with some of the drawings made by JMGB, highlighting the main Martian features.

## JOSÉ MARIA GONZALEZ
### DE BOGOTA

Encore un deuil pour notre Société ! L'ancien directeur de l'Observatoire national de Colombie, le fondateur, en 1880, de l'Observatoire Flammarion et de la Société scientifique Flammarion de Bogota, M. J.-M. GONZALEZ, apôtre passionné de la Science, qui faisait flotter le drapeau de la France dans cette belle cité du Soleil, sur l'équateur, à près de trois mille mètres d'altitude, vient de quitter cette terre, après avoir consacré, avec le plus rare désintéressement, sa vie tout entière à la cause sacrée du Progrès.

Il avait tenu à s'inscrire au nombre des membres fondateurs de la Société Astronomique de France, et notre *Bulletin* a plus d'une fois eu à enregistrer ses observations et ses travaux. Il y a deux ans encore, au sein d'une épouvantable guerre civile, il avait mis à la disposition de la France, sa « patrie intellectuelle », une mission d'ingénieurs spéciaux pour la continuation de la mesure de l'arc du méridien sur l'équateur. Président de l'Institut de Colombie, il était avec ses collègues, MM. Diodoro Sanchez, Enrique de Argaez, Indalecio Liévano et avec ses fils, ses dignes successeurs, à la tête du mouvement scientifique et de toutes les œuvres de progrès. Il n'a pas consacré moins d'un million à l'installation de son cher « Observatoire Flammarion ». C'était à la fois un esprit d'élite et un noble cœur.

La Société Astronomique de France prend une grande part à cette perte. Elle est sûre qu'entre les mains de ses collègues et de ses fils, l'œuvre si française de GONZALEZ ne périra pas, et elle adresse à sa famille ses sentiments de condoléance.

Figure 7: JMGB's obituary published in the Bulletin of the French Astronomical Society (1903) shortly after his death in 1903.

**Tablas**

Table 1. Orbital elements for Comet 1882 I assuming parabolic orbit, as presented by **González** (1882e) and **Vsekhsvyatskii** (1964), respectively.

| Orbital elements | González 1882(e) | Vsekhsvyatskii (1964) |
|---|---|---|
| Perihelion date | 10 June 1882 | 11 June 1882 |
| Longitude distance perihelion | 8,78367 | No data |
| Longitude of ascending node | 204º 54' 50'' | 206.94º |
| Inclination to the ecliptic plane | 73º 47' 30'' | 73.81º |
| Perihelion distance (q) | 2250000 leagues (10845000 km) | 0.0608 AU (9120000 km) |

Table 2. Parameters found by JMGB for comet Rordame-Quénisset 1893 II, assuming parabolic orbit.

| Orbital elements calculated by JMGB | Comet Rordame-Quénisset 1893 II |
|---|---|
| Perihelion date | 1893 July 7,291 |
| Argument of perihelion | 47º 7' 15.7'' |
| Longitude of ascending node | 337º 23' 25,9'' |
| Inclination to the ecliptic plane | 159º 58' 10.3'' |
| Perihelion distance (Log q) | 9,828936 |

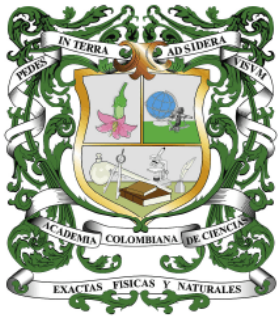

# Material suplementario

**The pioneering scientific endeavor and contributions of José María González Benito (1843-1903), the first Colombian modern astronomer**

**Los esfuerzos y aportes científicos de José María González Benito (1843-1903), el primer astrónomo moderno colombiano**

Freddy Moreno-Cárdenas, Santiago Vargas-Domínguez, Jorge Cuéllar-Moyano

**Autor correspondencia, correo electrónico**

## Contenido
- Texto
- Figuras 1S-10S
- Tablas 1S y 2S

Texto

## 5.2 Mars

Schiaparelli selected the four images shown in the left panel in Figure 6, which served to stress his ideas that there were narrow and dark areas on Mars, which, according to his hypothesis, were channels possibly built by an intelligent civilization on the red planet. Below we present the transcription of JMGB's report:

*The construction of the large building that today constitutes the Flammarion Observatory, was not completed in the most favorable time for observing Mars (August-September, 1892) and the new instruments are not yet installed, thus we had to limit ourselves to using the assembled instrument in the temporary installation, namely: an excellent Secretan equatorial telescope, 0m,108 and a Bardou telescope of 0m,095; however, we were able to successfully perform the strongest magnifications that these facilities, both due to the considerable altitude of our Observatory above sea level (2640 m) than from the great height of the planet above the horizon, from the clarity of the sky and the calm of the atmosphere for several nights" and complemented it with "convinced of the great difficulty that the observation of Mars represents and not having at the moment powerful enough devices to pretend to study the channels discussed and so interesting there, we limit ourselves to studying the most outstanding configurations and details: we present what we have really seen. On 1st August, at 10 p.m., under a magnificent sky, the look was admirable; the southern polar cap appeared in dazzling and still widespread light, embroidered with a rather accentuated dark line bearing a strongly marked notch at the 300° meridian. The boreal region was quite white like the outline of the star: the central parts showed a very pure matt white, and the dark regions were painted a very soft greenish-gray tone and slightly dark in the center.*

And follows

*From August 7 to 8, the southern cap showed in a sufficiently clear way the dark notch around the 300° meridian with a tendency to expand and that reached its maximum greatness from August 14 to 17.*

*The report from JMGB was followed by the letter signed by Schiaparelli (Figure 8S) commenting on the Colombian astronomer's drawings (Flammarion, 1909): "These observations are particularly interesting, given the altitude of this observatory established on the equator (4°35'48'' N). At the height of 2640 m, the atmosphere is very clear. Mr. Gonzalez is a careful and sincere observer. Of the 24 drawings that the wise founder of this equatorial establishment wanted to direct us, we chose four to be annexed here in our general documentation. Remarkably, the polar notch and the Main Sea (Lake Mœris) could be observed with the aid of a 108 mm. As for the decrease in the red coloration of the planet with its elevation in the sky, this may be due in part to an effect of our atmosphere that acts on the coloration of the Moon and the Sun, and in part to the objective of the lens, less achromatize, perhaps, by the blue and violet rays".*

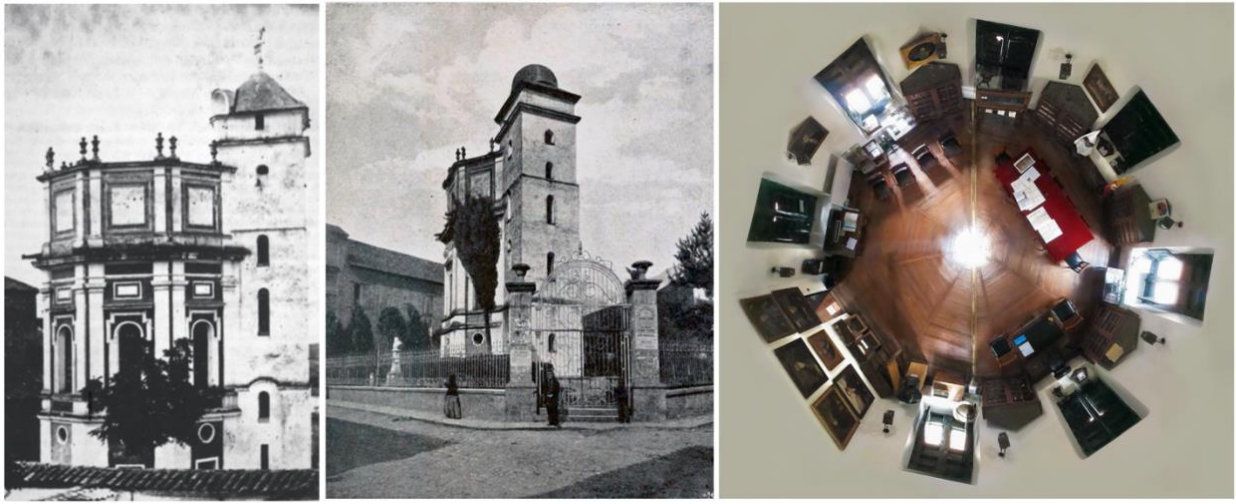

Figure 1S: Historical photos of the OAN. Left: In 1870 with the original configuration when JMGB assumed his first period as director of the institution. Middle: In 1898, with the new dome installed at the beginning of that decade under the direction of JMGB (repository of Biblioteca Nacional de Colombia). Right: In 2022, showing a top view of the main room with the meridian on the floor during the zenital day on April 1st at 12:00 m (local time). Photo taken by the authors.

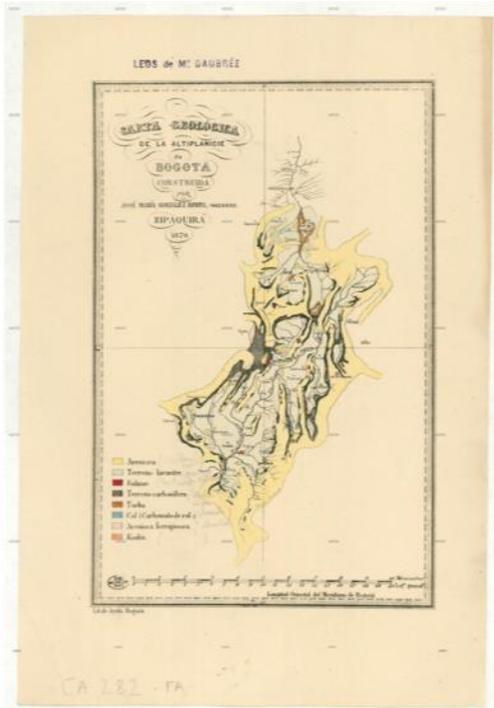

Figure 2S: Geological chart of the Bogotá plateau, published in 1870, which after 8 years of work by JMGB was donated by him to the National Government. With this work JMGB obtained the honorable mention in the outstanding Exhibition of the National Industry of 1871, the most important in the country (repository of the Biblioteca Nacional de Colombia).

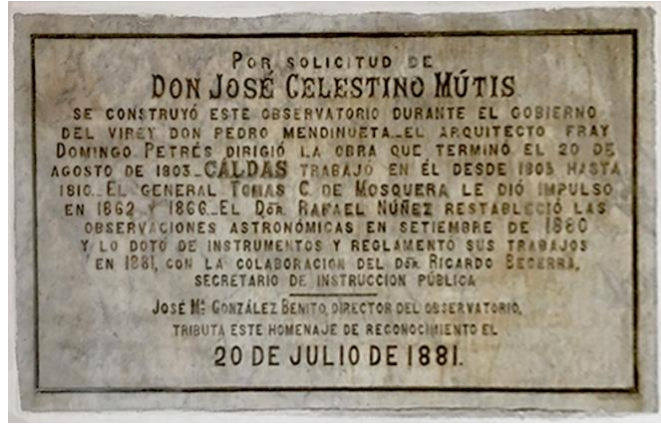

Figure 3S. Memorial plaque installed by JMGB in 1881 at the OAN to acknowledge the reinstatement of astronomical observations with the new instrumentation. Photo taken by the authors.

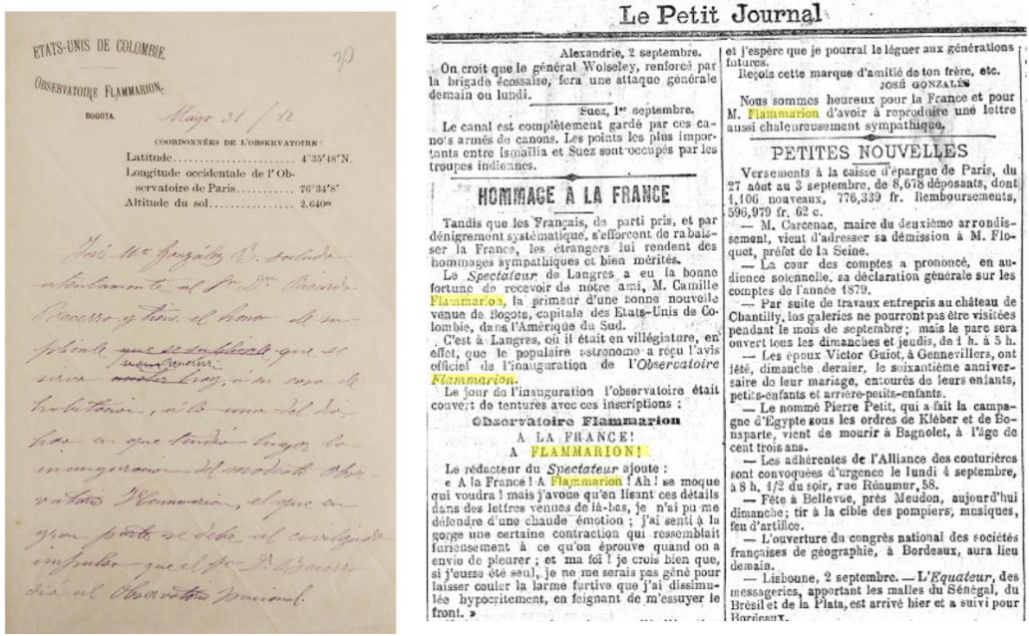

Figure 4S: Left: Invitation to the inauguration of the Flammarion Observatory in Bogotá, on May 31, 1882 (found in the repository of the Biblioteca Nacional de Colombia). Right: Publication in the LePetit Journal spotlighting the inauguration of the Flammarion Observatory honoring France.

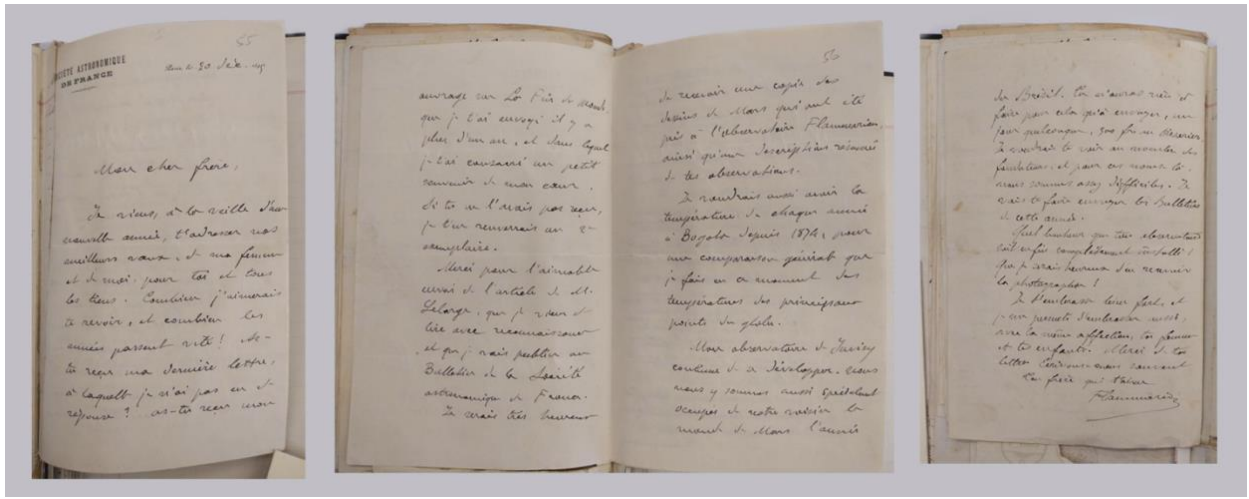

Figure 5S: Private communication sent by Flammarion to JMGB (**Flammarion**, 1895), found by the authors in the repository of Biblioteca Nacional de Colombia.

Proposed Observatory at Bogota, South America.

M. Gonzalez, Director of the National Observatory of Columbia, announced at the meeting that it was his intention to establish a Physical Astronomical Observatory at Bogotá, the capital of that State, at an altitude of about 3,000 mètres above the level of the sea, and in latitude 4° 30′ N. On account of the transparency of the atmosphere, M. Gonzalez believes that this observatory will be most favourably situated for delicate observations, such as the spectrum analysis of the heavenly bodies, especially of the Sun, the Zodiacal Light, &c. He intends to give up the direction of the National Observatory, so that he may be able to devote his whole attention, free from the control of the Government authorities, to this peculiar class of physical observation. M. Gonzalez expressed a desire that his private observatory might be considered as, in some measure, a dependence of this Society and the British Association, and he would therefore be happy to receive any suggestions from the leading Fellows as to the best means of utilising the observations which he hopes to make in such an exceptionally elevated locality. He is most desirous to carry out any recommendations he may receive so far as his resources will permit. M. Gonzalez will be assisted by his brother, as well as by a friend who is devotedly attached to the science.—[E. D.]

Figure 6S: JMGB proposal to the Royal Astronomical Society for the establishment of a mountain observatory in Bogotá at 3000 m.a.s.l, published in the Monthly Notices of the Royal Astronomical Society (**González**, 1874).

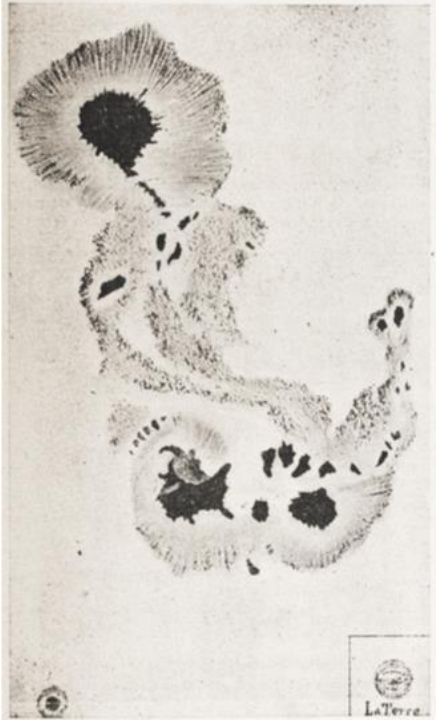

Figure 7S: Drawing of the great sunspot observed by JMGB in August 1893, corresponding to the growing phase of solar cycle 13, published in L'Astronomie the following year (Flammarion, 1894).

LETTRE DE M. SCHIAPARELLI. 77

Ces observations sont particulièrement intéressantes, étant donnée l'altitude de cet observatoire établi sur l'équateur (4°35′48″N). A cette hauteur de 2640 mètres, l'atmosphère est d'une heureuse limpidité. M. González est un observateur consciencieux et sincère. Sur les 24 dessins que le savant fondateur de cet établissement équatorial a bien voulu nous adresser, nous en avons choisi quatre pour être annexés ici à notre documentation générale. Il est remarquable que l'échancrure polaire et la mer Main (lac Mœris) aient pu être suivies à l'aide d'un 108$^{mm}$. Quant à la diminution de la coloration rouge de la planète avec son élévation dans le ciel, il est possible qu'elle soit due en partie à un effet de notre atmosphère — le même que celui qui agit sur les colorations de la lune et du soleil — et en partie à l'objectif, moins achromatisé peut-être pour les rayons bleus et violets (¹).

Figure 8S: Letter from the Italian astronomer Giovanni Schiaparelli published in Flammarion (1884), commenting on the detailed Mars observations carried out by JMGB.

vation spéciale, il n'a remarqué aucun essaim d'étoiles filantes pendant la nuit du 27, ni pendant la nuit précédente ni pendant la nuit suivante; mais, en revanche, un maximum tout à fait inattendu s'est présenté dans la nuit du 24. A partir de 8 heures du soir, voici les nombres d'étoiles filantes signalées:

| | |
|---|---|
| De 8ʰ à 9ʰ............ | 750 étoiles. |
| 9 à 10 ............ | 360 » |
| 10 à 11 ............ | 252 » |
| 11 à 12 ............ | 27 » |
| 12 à 1 ............ | 25 » |
| 1 à 2 ............ | 9 » |
| 2 à 3 ............ | 7 » |
| 3 à 4 ............ | 3 » |
| Total..... | 1433 étoiles. |

On voit que le maximum a dû avoir lieu avant 8 heures du soir. Le nombre total des étoiles filantes est beaucoup plus considérable que le total indiqué; car M. Gonzalès n'avait qu'un aide, de sorte qu'un grand nombre sont passées inaperçues. Le blanc prédominait, et l'on n'a remarqué que quelques rouges, bleues et jaunes. Ces étoiles ne venaient pas, comme celles du 27, de la constellation d'Andromède, mais de celle du Lion: une carte, que M. Gonzalès nous a remise, place le radiant au nord de Régulus, près de ζ du Lion. La plupart se dirigeaient du sud-ouest vers le nord-est. Cet essaim n'appartiendrait-il pas au système des météorites du 14 novembre? Le fait est d'autant plus probable que l'inclinaison de cet essaim et de la comète Tempel de 1866, sur le plan de l'orbite terrestre, n'est que de 18 degrés.

Quoiqu'il en soit de cette pluie du 24, celle du 27 a

Figure 9S: Report from JMGB on the unknown shooting stars on the night of November 24, 1872, whose origin was the constellation of Leo published in Flammarion (1874).

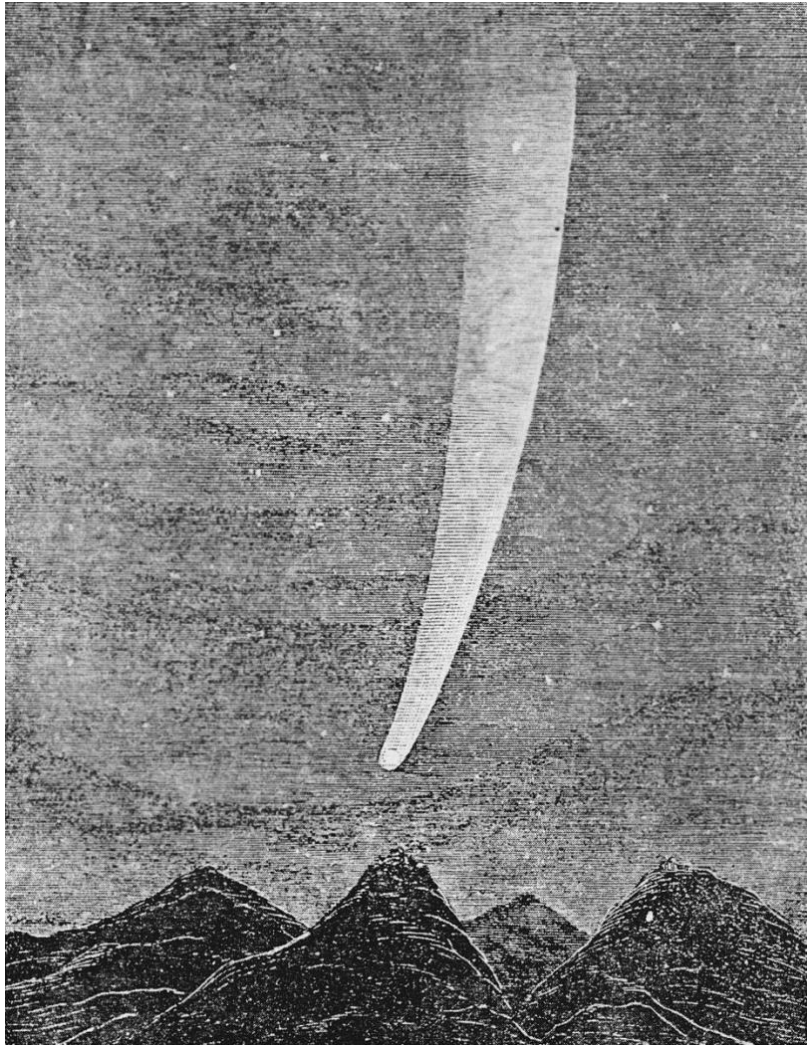

Figure 10S: Engraving of the Great Comet of 1882 made by Alberto Urdaneta as it was observed in Bogotá in September of that year. JMGB sent the report of his observation to the French Astronomical Society (Papel Periódico Ilustrado, 1882).

Table 1S. Ephemeris with the information about the time contacts for the transit of Venus on December 6, 1882, as presented in González (1882c).

| Transit Phase | Time |
|---|---|
| Contact I (ingress exterior) | 9h 3m 18,07s |
| Contact II (ingress interior) | 9h 23m 33,15s |
| Central phase | 12h 13m 2,06s |
| Contact III (egress interior) | 3h 2m 30,97s |
| Contact IV (egress exterior) | 3h 22m 52,14s |

Table 2S. Chronology of the scientific life of the Colombian astronomer José María González Benito (JMGB).

| Year | Event |
|---|---|
| 1803 | NAO was founded as part of the Botanical Expedition of the New Kingdom of Granada. |
| 1843 | JMGB was born in Zipaquirá on September 1. |
| 1858 | JMGB completes studies at the college and continues to receive private mathematics classes. He accompanies the engineer Manuel Ponce de León, appointed by the Government to draw up the plans of the lands of the Zipaquirá, Nemocón, Tausa and Sesquilé to make the demarcation of salt flats |
| 1862 | JMGB completes his private studies acquiring skills to raise topographic maps.<br>JMGB is appointed as assistant to the engineer Indalecio Liévano Reyes, director of the OAN, in the design of railway lines.<br>The study of the geology of Colombia begins, collecting various minerals and fossils in expeditions to study geological formations in various regions of the country.<br>Redesign of the meridian at the NAO. |
| 1863 | On December 31st, JMGB's position as an assistant in the OAN culminates. |
| 1864 | JMGB is hired to carry out topographic surveys in the region of Santander.<br>JMGB's first trip to Europe. He enrolls as a student at the Central School and attends other courses at the Sorbonne, mainly geology and astronomy. He meets Joseph Alfred Serret, Pouissex, Urbain Le Verrier, Yvon Villarcean, Jean Baptiste Élie de Beaumont, Jean-Baptiste Boussingault, Adolphe Brongniart. He is very enthusiastic about mineralogy and astronomy, acquiring books, instruments, and an abundant geological and paleontological collection. |
| 1866 | JMGB returns to Colombia. He is appointed as assistant to the Central Office of the Brigade of National Engineers, and as head of the OAN observations diary.<br>JMGB obtains the degree of Engineer. |

| Year | Event |
|------|-------|
| 1867 | JMGB studies with Liévano the Leonid meteor shower. Creation of the Universidad Nacional de Colombiaa. |
| 1868 | JMGB is appointed professor of Meteorology and Astronomy at the Universidad Nacional de Colombia, receiving the position of Director of the Astronomical Observatory (first term).<br>JMGB returns to Zipaquirá to finish a geographical chart of the savannah and the highlands of Bogotá. |
| 1870 | JMGB publishes the "Geological Chart of the Plateau of Bogotá", which compiles a work that had begun more than eight years before, and which he offers as a donation to the Government. |
| 1871 | JMGB returns to Bogotá for appointment as professor of Geology and Paleontology at the Universidad Nacional de Colombia, being a pioneer in the training of students in these areas in the country.<br>JMGB is appointed director of the OAN (second time) and principal professor of astronomy, opening in the country studies at the university level in this area, as he had done with geology and paleontology.<br>Publication of the meteorological and astronomical observations made at the OAN in the Annals of the Universidad Nacional de Colombia.<br>Realization of the first National Exhibition, where JMGB contributes with a copious collection of rocks, fossils and minerals of the country, together with the geological map of Cundinamarca that he had made.<br>JMGB receives a Diploma of Honor for the geological map of the Sabana de Bogotá.<br>JMGB is named as a member of the Academia de Ciencias Naturales, created in 1868 at the Universidad Nacional de Colombia. |
| 1872 | JMGB retirees from the Directorate of the OAN.<br>JMGB receives membership of the Society of Light, part of the Institute of Arts and Crafts, which aims to spread useful knowledge for the country, supporting as a professor of geology courses for the general public.<br>JMGB is appointed for the third time as Director of the NAO.<br>The OAN is temporarily closed. |
| 1873 | JMGB is appointed by decree, and for the fourth time, director of the OAN, and professor of astronomy and geodesy at the Escuela de Ingeniería, Universidad Nacional de Colombia. |
| 1874 | JMGB travels to England. He proposes the construction of a high mountain astronomical observatory in Colombia, through a publication in the journal Monthly Notices of the Royal Astronomical Society. JMGB is appointed member of the Royal Astronomical Society. |
| 1875 | JMGB returns to Colombia and acquires astronomical instruments for the OAN. |
| 1877 | Report of the shooting stars in Andromeda observed on November 24, 1876 and published in the eighth volume of the "Studies and readings on astronomy" of 1877 at the Gauthier-Villars establishment, Paris. |
| 1878 | JMGB marries Maria Danies Kennedy in the city of Rioacha, at the north of Colombia.<br>JMGB's second trip to Europe. |
| 1879 | JMGB returns to Colombia. Foundation of the Flammarion Observatory in Zipaquirá |
| 1880 | JMGB receives the shipment from Mr. Secretan consisting of an equatorial telescope (number 417 of the 1874 Catalog).<br>Creation of the Flammarion Scientific Society, corresponding member of the French Astronomical Society.<br>JMGB is appointed for the fifth time as director of the NAO. Return to Bogotá. The physical and instrumental renovation of the OAN begins. |
| 1881 | Installation of the NAO dome, 16 cm refractor, complete weather station. The NAO is designated an area for systematic observation between declination 40 and 55 north and participates in the unification of time management. |
| 1882 | Inauguration of the Flammarion Observatory in Bogotá, in the Parque de Los Mártires, with the invitation of prominent personalities, such as the Ambassador of France, and that of Chile. The instrumentation included 12" and 8" telescopes. JMGB creates the publication "Annals of the National Astronomical Observatory of Colombia", intended to publish the works of the establishment, which completed six issues.<br>Observation of comet 1882I on June 10.<br>Observation of the Great September Comet on August 14 (15 days before the official sighting). |
| 1883 | Creation of the Juvisy-Sur-Orge Observatory by Flammarion in France. |
| 1884 | JMGB is invited to the World Congress in Washington where he delegates his participation. The zero meridian is adopted at Greenwich. Unification of the hour.<br>The Flammarion Observatory moves to the three-story house number 340 on Carrera 7, former Calle de la Carrera, which will be a temporary headquarters. |

| 1891 | JMGB surrenders for good the position of director of the OAN. |
| --- | --- |
| 1892 | Construction of the definitive headquarters of the Flammarion Observatory begins, in building number 90 on 16th Street.<br>Observation of the planet Mars. |
| 1893 | JMGB is presented by Camille Flammarion and Anatole Bouquet de La Grye to the Astronomical Society of France as a founding member.<br>JMGB is co-founder of the Institute of Arts and Crafts of Bogotá, which sought to teach non-formal education to low-income students in the capital.<br>Publication of the solar observations made at the Flammarion Observatory, in L'Astronomie, the great sunspot of August 1893.<br>Observation of comet Rordame-Quénisset on July 7, 1893, before the official discovery on July 8, 1893. |
| 1895 | Relocation of the Flammarion Observatory to its definitive headquarters. |
| 1896 | Completion of the construction of the Flammarion Observatory: North latitude: 4º 36' 43", Greenwich longitude: 74º 131' 33", altitude of the ground floor: 2645 m., altitude on the roof: 2659 m.<br>JMGB visits the eminent architect Gaston Lelarge to the Flammarion Observatory and publishes an article describing his visit. |
| 1899 | JMGB establishes links with the French Astronomical Society to participate in the project to measure the arc of the meridian at the equator. Studies the meteor shower of the Leonids. |
| 1903 | JMGB proposes the creation of the Institute of Colombia, bringing together the academies of mathematics, natural sciences, and moral and political sciences.<br>JMGB died in Bogotá on July 28, the day before the inauguration of the Institute of Colombia that was planned to be held in the Colon Theater. |